\documentclass[aps,prl,superscriptaddress]{revtex4-2}

\usepackage{graphicx}% Include figure files
\usepackage{dcolumn}% Align table columns on decimal point
\usepackage{bm}% bold math
\usepackage[utf8]{inputenc}
\usepackage[T1]{fontenc}
\usepackage{mathptmx}
\usepackage{etoolbox}
\usepackage{braket}
\usepackage{hyperref}
\usepackage{color}
\usepackage{subfiles}
\usepackage[svgnames]{xcolor}
\usepackage{ulem}
\usepackage{subcaption}
\usepackage{babel}
\usepackage[symbol]{footmisc}

\begin{document}

%\preprint{}

\title{Near-Infrared and Telecommunication-Wavelength Photon-Pair Source in Optical Fiber%Highly Non-Degenerate Fiber-Based Source of Multiplexed Photon Pairs
}


\author{Keshav Kapoor\footnotemark[1]\footnotetext{Indicates that these authors contributed equally to this work.} \footnotemark[2]\footnotetext{kkapoor2@illinois.edu}}
\affiliation{Department of Physics, University of Illinois Urbana-Champaign, Urbana, Illinois 61801, USA}
\affiliation{Illinois Quantum Information Science \& Technology Center (IQUIST), University of Illinois Urbana-Champaign, Urbana, Illinois 61801, USA}
\author{Dong Beom Kim\footnotemark[1]}
\affiliation{Department of Physics, University of Illinois Urbana-Champaign, Urbana, Illinois 61801, USA}
\affiliation{Illinois Quantum Information Science \& Technology Center (IQUIST), University of Illinois Urbana-Champaign, Urbana, Illinois 61801, USA}
\affiliation{School of Applied and Engineering Physics, Cornell University, Ithaca, NY 14853, USA.}
\author{Kriti Shetty}
\author{Virginia O. Lorenz}
%\thanks{}
\affiliation{Department of Physics, University of Illinois Urbana-Champaign, Urbana, Illinois 61801, USA}
\affiliation{Illinois Quantum Information Science \& Technology Center (IQUIST), University of Illinois Urbana-Champaign, Urbana, Illinois 61801, USA}


\date{\today}

\begin{abstract}
%Real-world quantum networks require bright, multiplexed photon sources for high-speed, versatile communication. The current optical-fiber infrastructure merits the fiber-based entangled sources that enable advanced quantum networking protocols.
We present a photon-pair source in commercially available optical fiber that produces paired photons at telecommunication and near-infrared (NIR) wavelengths. The highly nondegenerate pairs are 700 nm apart: one in the 1500 nm E- and S-band telecommunication range and the other in the 830 nm NIR range. The high non-degeneracy means the photon pairs are far-detuned from Raman noise, resulting in a high coincidence-to-accidental ratio even while operating at room temperature. The source produces two spectrally and spatially distinct phase-matched processes with low spectral cross-talk, distinct transverse spatial modes in the NIR, and a single fundamental spatial mode in the telecommunication range. The source's room-temperature operation, off-the-shelf materials, and multiplexing potential make it promising for deployment in quantum networks.
%The source is bright % (32.5 kcps) 
\end{abstract}

\maketitle


\section{Introduction}

Fiber-based sources of photon pairs have been shown to be highly compatible with existing fiber-based networking infrastructure \cite{Mueller:24, Valivarthi_Davis_Pena_Xie_Lauk_Narvaez_Allmaras_Beyer_Gim_Hussein_et_al._2020, Alshowkan:22, li2005optical, Liu_Liu_Jin_Lin_Li_You_Feng_Liu_Cui_Zhang_et_al._2024, chakraborty2025towards, Wang_Hong_Friberg_2001, fiorentino2002all}, towards the building of quantum networks \cite{Chung_Eastman_Kanter_Kapoor_Lauk_Pena_Plunkett_Sinclair_Thomas_Valivarthi_et_al._2022, Wehner_Elkouss_Hanson_2018, Azuma_Economou_Elkouss_Hilaire_Jiang_Lo_Tzitrin_202,Pan_Bouwmeester_Weinfurter_Zeilinger_1998,Liao_Cai_Liu_Zhang_Li_Ren_Yin_Shen_Cao_Li_et_al._2017,Davis_Valivarthi_Cameron_Pena_Xie_Narvaez_Lauk_Li_Taylor_Youssef_et_al._2025, Pompili_Hermans_Baier_Beukers_Humphreys_Schouten_Vermeulen_Tiggelman_Dos_Santos_Martins_Dirkse_et_al._2021, van_Leent_Bock_Fertig_Garthoff_Eppelt_Zhou_Malik_Seubert_Bauer_Rosenfeld_et_al._2022, Jiang_Chen_Yan_Lu_Wen_An_Chen_Liu_Liu_Xie_et_al._2025, Wengerowsky_Joshi_Steinlechner_Hubel_Ursin_2018, Kapoor_Xie_Chung_Valivarthi_Pena_Narvaez_Sinclair_Allmaras_Beyer_Davis_et_al._2023, Fan_Luo_Guo_Wu_Zeng_Deng_Wang_Song_Wang_You_et_al._2025}.
They benefit from the ready availability and customizability of materials, a wide range of co-linear phase matching conditions, and mode-matched coupling into existing fiber-based networking infrastructure~\cite{Garay-Palmett:23}.
For useful quantum applications, the sources need to be be high rate and low noise while having at least one photon at telecommunication wavelengths to be compatible with existing infrastructure.
It is also advantageous to have nondegenerate photon pairs at visible/near-infrared (NIR) and telecommunication wavelengths to bridge the gap between fiber-based and free-space networks~\cite{Oh_Jennewein_2024, Liu:25, sundaram2025heralded}. 
Additionally, the ability to have a multiplexed source can further improve the success rates of quantum protocols~\cite{Sinclair_Saglamyurek_Mallahzadeh_Slater_George_Ricken_Hedges_Oblak_Simon_Sohler_et_al._2014, Chakraborty_Das_van_Brug_Pietx-Casas_Wang_Amaral_Tchebotareva_Tittel_2025, Ruskuc_Wu_Green_Hermans_Pajak_Choi_Faraon_2025}.
Fiber-based sources that have addressed some of these benchmarks include those that are based on dispersion-shifted fiber producing slightly nondegenerate photon pairs at telecommunication wavelengths~\cite{li2004all}, with liquid nitrogen cooling to suppress Raman noise~\cite{Raman_Krishnan_1928}, and ring-core fiber producing highly nondegenerate photons at telecommunication and visible wavelengths~\cite{Liu:25}, in which the photons are produced in annular orbital-angular-momentum modes.

Here, we present a source of highly nondegenerate photon pairs in the telecommunication and NIR wavelengths based on spontaneous four-wave mixing (SFWM) in commercially available telecommunication polarization-maintaining fiber (PMF). The birefringent phase-matching of the PMF leads to highly nondegenerate photon pairs that are far detuned from the pump, avoiding substantial Raman noise and the need for cooling the fiber~\cite{Smith:09}.
Additionally, with an optimal choice of pump wavelength given the telecommunication fiber core size, we see multiple SFWM processes due to the PMF supporting multiple spatial modes~\cite{Kim:25} for the pump and NIR photon wavelengths. Simultaneously, the telecommunication photons are produced in the fundamental spatial mode, which makes the source compatible with commercially deployed single-mode fiber (SMF). This opens possibilities for spatio-spectral multiplexing.
In this paper, we use joint spectral intensities, coincidence count rates, and cross-correlation measurements to characterize our photon-pair source~\cite{Fox_2013}.
From these characterizations we show that this source is suitable for deployment in quantum networks for free-space and in-fiber distribution at high rates, with the potential for spectral multiplexing. 

\section{Results}
The SFWM process relies on the $\chi^{(3)}$ nonlinear optical susceptibility, which occurs in centro-symmetric materials such as optical fibers made out of fused silica \cite{Wang_Hong_Friberg_2001}. 
Birefringent fiber, such as PMF \cite{Smith:09}, allows phase-matching conditions that support the creation of highly nondegenerate daughter photons. We use the cross-polarized scheme, in which the pump is launched along the slow axis to create daughter photons along the fast axis \cite{Garay-Palmett_Cruz-Delgado_Dominguez-Serna_Ortiz-Ricardo_Monroy-Ruz_Cruz-Ramirez_Ramirez-Alarcon_URen_2016}.
This process also yields diverse spectral and spatial phase-matching conditions we can explore~\cite{Kim:25}. 
In particular, the fiber supports higher-order spatial modes of both the pump and signal photons and thus can produce daughter photons which are spectrally distinct from each other.

\subsection{Experimental Setup}
\begin{figure}
    \includegraphics[width=\linewidth]{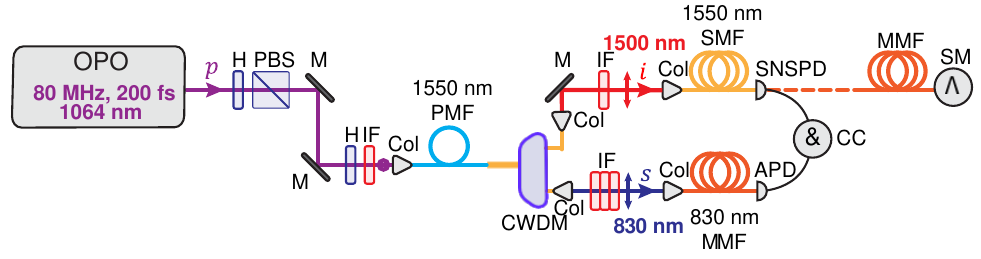}
    \caption{Experimental setup for generating and characterizing telecommunication-NIR photon pairs created via SFWM. The 1-meter-long 1550 nm PANDA PMF is pumped with 1064 nm fs-pulsed light from an OPO. The polarization is controlled by half waveplates (H) and a polarizing beam splitter (PBS) before spectral filtering with an interference filter (IF) and then coupling into the PMF via a collimation package (Col). The PMF produces photon pairs at 1490 nm and 830 nm and 1430 nm and 850 nm. After the PMF, the signal and idler photons are separated using a coarse wavelength division multiplexer (CWDM). They are further spectrally filtered in free space before being collected back into SMF or multi-mode fiber (MMF), and detected on the superconducting nanowire single-photon detectors (SNSPDs), avalanche photodiodes (APDs) or a spectrometer (SM). The coincidences are measured using a coincidence counter (CC).}
    \label{fig:experimental_setup}
\end{figure}

Our experimental setup is illustrated in Fig.~\ref{fig:experimental_setup}. 
We use a one-meter-long telecommunication PMF (PANDA 1550, Thorlabs) as the SFWM medium.
The pump beam is produced by an optical parametric oscillator (OPO, Inspire) at 1064 nm with a spectral bandwidth of 6~nm, a pulse duration of about 200~fs, and a repetition rate of 80~MHz. Half-wave plates and a polarizing beam splitter align the polarization of the pump such that it coincides with the slow axis of the PMF. Photons pairs are produced on the fast axis at signal (idler) wavelengths of 830 (1490) nm and 850 (1430) nm, which we label as process 1 and 2, respectively.  
The PMF is connected to a coarse wavelength division multiplexer (CWDM, Oz Optics) to separate the NIR and telecommunication daughter photons.
The daughter photons are sent to their respective free-space paths to be further filtered using interference filters to remove the residual pump.

\subsection{Spectrum and JSI}
 To obtain the marginal spectra and joint spectral intensity (JSI) of the daughter photons, we utilize single-photon-level spectrometers (Andor SR303i, SR500i Oxford Instruments) and a multi-mode fiber to collect all the spatial modes of the NIR photons and their associated spectra.
Two spectral features at signal wavelengths arise from the pair-generation process, one at 830 nm and the other at 850 nm, each corresponding to separate SFWM processes (see Supplemental Section 1).

\begin{figure}[htbp]
      \includegraphics[width=\linewidth]{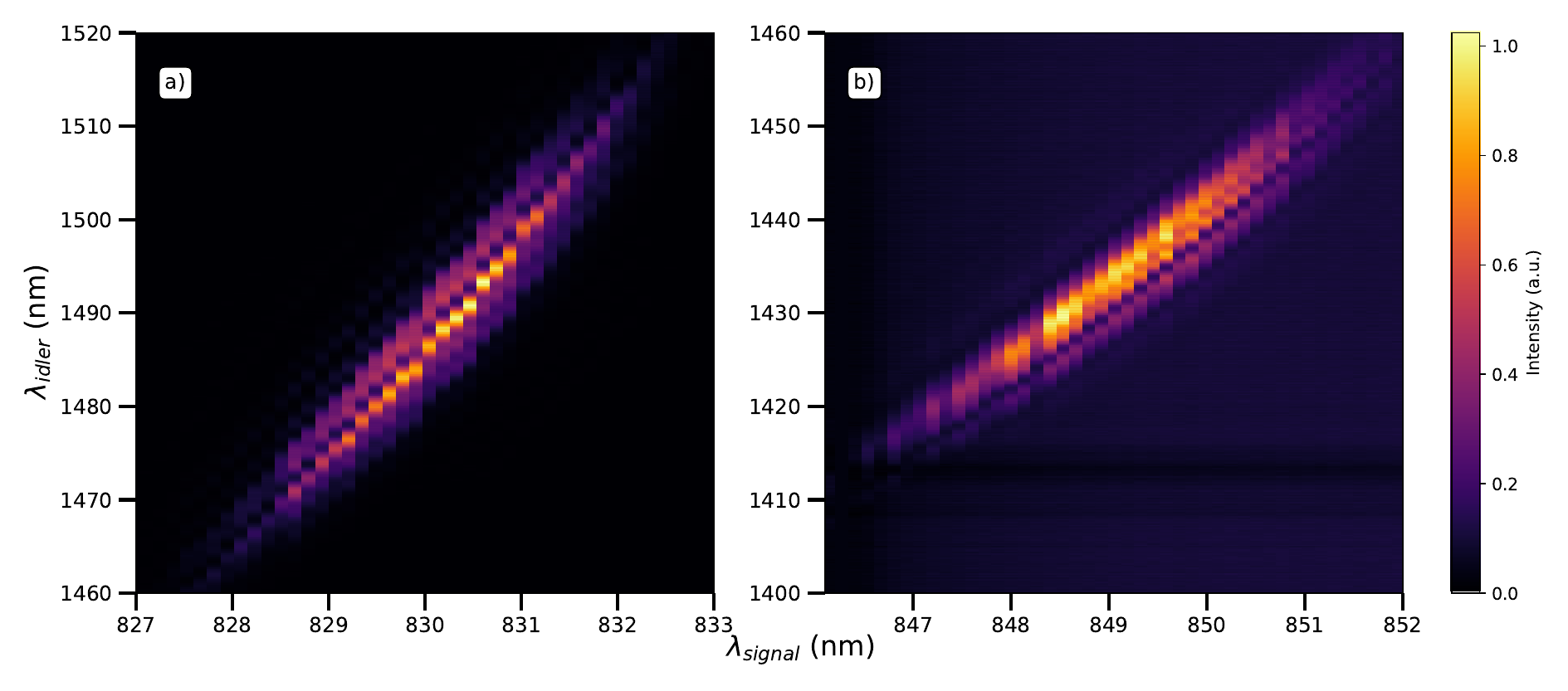}
     \caption{Joint spectral intensities (JSIs) of the photon pairs from stimulated four-wave mixing in the polarization-maintaining fiber. We obtain this JSI via stimulated emission through a tunable seed laser centered at (a) 830 nm for processs 1 and (b) 850 nm for process 2. The intensities are independently normalized to the maximum value for each process.}
     \label{fig:JSI}
\end{figure}

We use stimulated emission of the four-wave mixing (FWM) process for fast and high-resolution measurements of the JSI~\cite{Fang:16}. 
The stimulated telecommunication photons are collected with a single-mode telecommunication fiber.
We use a tunable CW Ti:Sapph laser (Coherent 899 Ring) as a NIR-wavelength seed to stimulate the photon-pair generation process.
We then scan the seed laser wavelengths to obtain the JSI, which are shown in Fig.~\ref{fig:JSI}.
The diagonal streaks in the JSI intensity next to the main intensity peak are likely caused by the chirp in the pump pulse as it travels through the PMF~\cite{Kim:25, Agrawal_2019}. We observe process 1 and process 2, which we experimentally verify corresponding to different spatial modes supported in the PMF, as described in Supplemental section 2. The JSIs in Fig.~\ref{fig:JSI} are normalized to the maximum intensity for each process independently, making them appear similar in brightness; as we shall see in the following rate measurement results, process 1 (a) is stronger than process 2 (b). From the data, we note that the process 1 telecommunication-wavelength photons are in the telecommunications E-band and the process 2 telecommunication-wavelength photons are in the telecommunications S-band.

\subsection{Rates and CAR}
\begin{figure}[ht]
    \centering
    \includegraphics[width=0.5\linewidth]{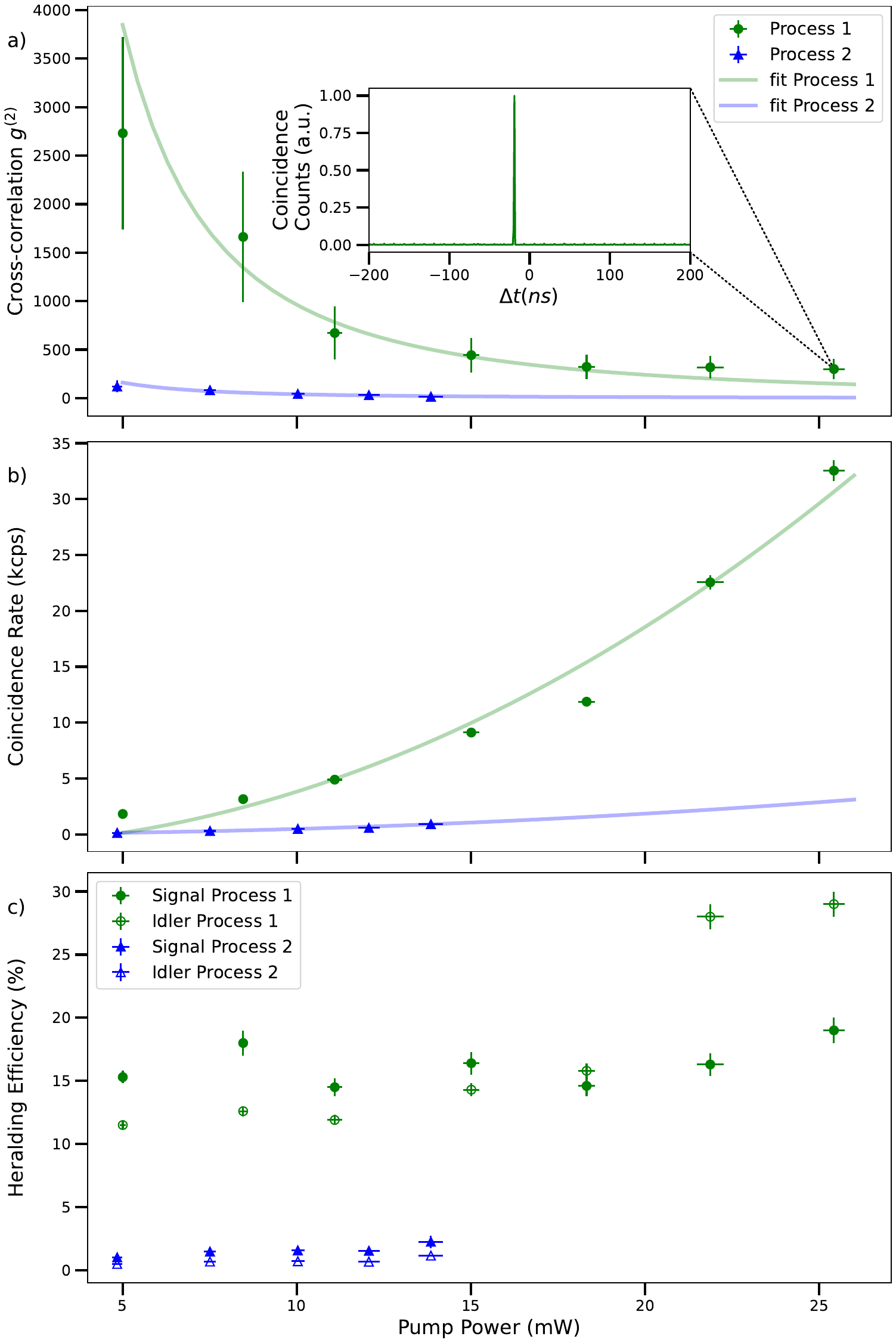}
    \caption{Photon-pair statistics as a function of average pump power relative to the signal and idler for processes 1 and 2. (a) Cross-correlation $g^{(2)}$ of the photon pairs. The inset shows the coincidence-count histogram used to calculate the $g^{(2)}$ for process 1 at the maximum pump power. (b) Coincidence count rate.  (c) Heralding efficiency. For a given pump power, process 1 produces higher rates and heralding efficiency than process 2.}
\label{fig:power_g2_rates}
\end{figure}

To measure the singles count rates, coincidence count rates, and cross-correlation $g^{(2)}$ of the source, we spectrally filter the NIR and telecommunication photons to distinguish and herald the photons generated from different SFWM processes. After filtering, we send them to single-photon detectors, with superconducting nanowire single-photon detectors (SNSPDs, Quantum Opus) used for the telecommunication photons and avalanche photodiodes (APDs, Excelitas SPCM-AQ4C) for the NIR photons.
The SNSPDs have an efficiency over 90 \% while the APDs have an efficiency of 45 \% at the corresponding wavelengths.
We calculate the cross-correlation $g^{(2)}$ and associated coincidence-to-accidental ratio (CAR) from the coincidence histogram as $g^{(2)} = \frac{N_c}{N_a}$, where $N_c$ is the coincidence counts at time zero and $N_a$ are the average over the accidental counts that occur periodically at the inverse repetition rate of the pump laser (every 12.5 ns); see raw histogram data in the inset of Fig.~\ref{fig:power_g2_rates}(a).
The measured coincidence count rate, cross-correlation $g^{(2)}$, and heralding efficiency as a function of pump power are plotted in Fig.~\ref{fig:power_g2_rates}.
We do not have data above 15 mW for process 2 as that is the maximum power that could be coupled to the spatial mode associated with process 2. 

Single and coincidence counts data were taken with a 2 ns coincidence window and 10 minute integration time. For process 1, at a pump power of about 25 mW,  we see singles rates of 175 $\pm$ 8 kcps and 112 $\pm$ 2 kcps for the NIR and telecommunication photons, respectively.
We measure a coincidence count rate of $32.5 \pm 0.9$ kcps, and a $g^{(2)}$ of $300 \pm 100$.
For process 2, at a pump power of 14 mW we see singles rates of 40.2 $\pm$ 0.8 kcps and 79 $\pm$ 2 kcps for the NIR and telecommunication photons, respectively.
We measure a coincidence count rate of $910 \pm 6$ cps, and a $g^{(2)}$ of $15 \pm 5$. We assume Poissonian statistics and perform error propagation to calculate the uncertainties. We limit the maximum PMF-coupled pump power to about 25 mW to avoid multi-pair generation. As seen in Fig.~\ref{fig:power_g2_rates}, process 1 shows higher singles and coincidence rates as well as cross-correlation $g^{(2)}$. The difference in rates between the two processes can be mainly attributed to the difference in spatial overlap integral, or the degree to which the participating fields in the SFWM process interact spatially~\cite{Kim:25}.

\section{Discussion}

Rates are impacted by loss and detector efficiency, with the 800 nm photons experiencing excess loss in components designed for telecommunications wavelengths, such as the CWDM and PMF. The 800 nm photons are also detected less efficiently by the APDs compared to the SNSPDs. Accounting for transmission losses and detector efficiencies, the coincidence rate at the output of the source is estimated to be well over 100 kcps.
See Supplement section 4 for more details on the losses and efficiencies. Currently, we are not actively controlling the spatial mode into the PMF and therefore are not controlling the relative strength of the two SFWM processes.
This can be done by incorporating a phase mask or spatial-light modulator in the pump arm \cite{Gissibl_Schmid_Giessen_2016, Savage_2009, Kim:25}.
This would potentially allow higher rates for process 2, which currently are significantly lower than process 1, or vice versa to further optimize the rates for process 1.

\section{Conclusion}
We have shown a high-coincidence-rate, high-CAR fiber-based source of photon pairs with two distinct spatial-mode SFWM processes. 
The spatial modes of the NIR signal photons from processes 1 and 2 can be spectrally isolated. 
The idler photons from processes 1 and 2 are at telecommunication wavelengths that can be spectrally separated while having a single spatial mode that is compatible with fiber transmission.
These properties make the source suitable for applications in multiplexed quantum repeater networks. 

The spectrum of the daughter photons can be readily modified by changing fiber length, fiber type, and the pump center wavelength \cite{Garay-Palmett:23, Smith:09},  making this a versatile source for quantum networking applications.  
The off-the-shelf components and ease of assembly also make it cost-effective and accessible for expanding quantum networks in regions that do not typically have direct access to quantum optical equipment, for example, as involved in the Public Quantum Network \cite{kapoor2025public}. 

In the future, we aim to use this source to generate polarization entanglement using cross-spliced PMFs or a Sagnac-interferometer \cite{Meyer-Scott:13, Fang:14}. 
From our JSI, we can see that the photons from such an entangled pair source source can be separated in two ways:
first, coarsely based on their inter-process spectral separations using CWDMs; and second, as shown recently \cite{Mueller:24, Alshowkan:22}, based on their intra-process spectral correlations using dense wavelength division multiplexers (DWDM).
These demultiplexing configurations can support a myriad of possible network configurations, such as mesh- and hub-based quantum network models, as well as other network and sensing protocols such as multiplexed quantum repeaters~\cite{Cao_Yu_Cai_2013,Gupta_Agarwal_Mogiligidda_Kumar_Krishnan_Chennuri_Aggarwal_Hoodati_Cooper_Ranjan_Bilal_Sheik_et_al._2024, Sinclair_Saglamyurek_Mallahzadeh_Slater_George_Ricken_Hedges_Oblak_Simon_Sohler_et_al._2014}, 
quantum telescopy, spectroscopy, and spatial-mode sensing \cite{Stas_Wei_Sirotin_Huan_Yazlar_Arias_Knyazev_Baranes_Machielse_Grandi_et_al._2025,Gottesman_Jennewein_Croke_2012, Brown_Allgaier_Thiel_Monnier_Raymer_Smith_2023, Diaz_Zhang_Lorenz_Kwiat_2021,Kira_Koch_Smith_Hunter_Cundiff_2011,Choi_Pluchar_He_Guha_Wilson_2024}. 

\section{Acknowledgments}
We would like to thank Paul G. Kwiat, Prem Kumar, and Fermilab collaborators Cristián Peña, Si Xie, and Andrew Cameron for useful discussions. 
This work was supported by the DOE Grant No. 712869, "Advanced Quantum Networks for Science Discovery,” and the NSF Award No. 2016136, NSF QLCI HQAN.
Data underlying the results presented in this paper are not publicly available at this time but may be obtained from the authors upon reasonable request. 
The authors declare no conflicts of interest.

See Supplement for supporting content.

\bibliography{bib_telecomsfwm}

\maketitle

\section{Near-Infrared Spectrum}
\begin{figure}[htbp]
      \centering
      \includegraphics[width=0.5\linewidth]{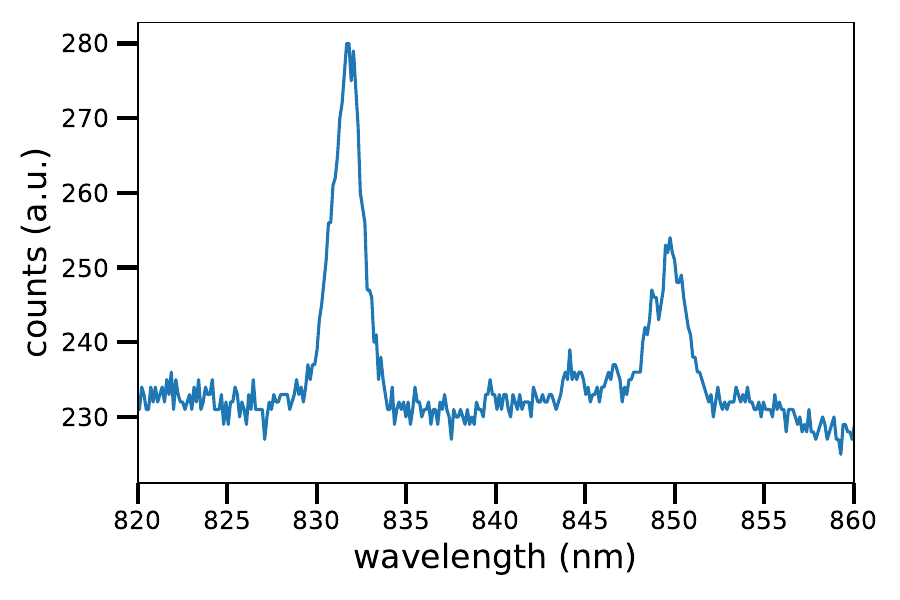}
     \caption{Example of the signal (NIR) spectrum we obtain from the PMF.}
     \label{fig:signal_spect}
\end{figure}

We use the Andor 303i to measure the spectrum of the near-infrared (NIR) photons from the 1-meter-PMF we are using, as mentioned in the Spectrum and JSI section of the main text. 
Here we can see the spectrum of the (NIR) photons created in the PMF.
We collect them into a MMF and measure the signal peaks from both processes.
This is shown in Fig. \ref{fig:signal_spect}.

%\section{Cross-correlation measurements as a function of pump power}
%We take cross-correlation $g^{(2)}$ measurements at multiple pump powers for both processes. 
%In the main manuscript we show these values plotted versus power. 
%Here we show the coincidence histograms for those different powers as shown in Figs. \ref{fig:g2_p1_25mW,fig:g2_p1,fig:g2_p2}.

%\begin{figure}[htbp]
%      \includegraphics[width=\linewidth]{figures/figures_supplemental/25mW_fund_20251008.pdf}
%     \caption{$g^{(2)}$ plot for process one at the maximum input pump power of about 25 mW.}
%     \label{fig:g2_p1_25mW}
%\end{figure}

%\begin{figure}[htbp]
%      \includegraphics[width=\linewidth]{figures/figures_supplemental/all_fund_20251008.pdf}
%     \caption{$g^{(2)}$ plots for process 1 at pump powers of about (a) 22 mW (b) 18 mW (c) 15 mW (d) 12 mW (e) 8 mW and (f) 5 mW. Small accidental counts peaks on the sides become more prominent with higher pump powers.}
%     \label{fig:g2_p1}
%\end{figure}

%\begin{figure}[htbp]
%      \includegraphics[width=\linewidth]{figures/figures_supplemental/all_higher_20251010.pdf}
%     \caption{$g^{(2)}$ plots for process 2 at pump powers of about (a) 14 mW (b) 12 mW (c) 10 mW (d) 7 mW and (e) 5 mW.}
%     \label{fig:g2_p2}
%\end{figure}

\section{Spatial Mode Dependence of Processes} \label{supp:spatial_mode}
To confirm that the process 2 is associated with a higher-order spatial mode, we utilize stimulated emission like we do with the JSI.
We set the seed wavelength to either the process 1 and 2 based off the spectrum we obtained for the NIR signal earlier.
We then control the spatial mode into the PMF by changing the coupling into the fiber and utilizing a camera at the output of the PMF to confirm the spatial mode that propagates through the fiber. 

\begin{figure}[htbp]
      \includegraphics[width=\linewidth]{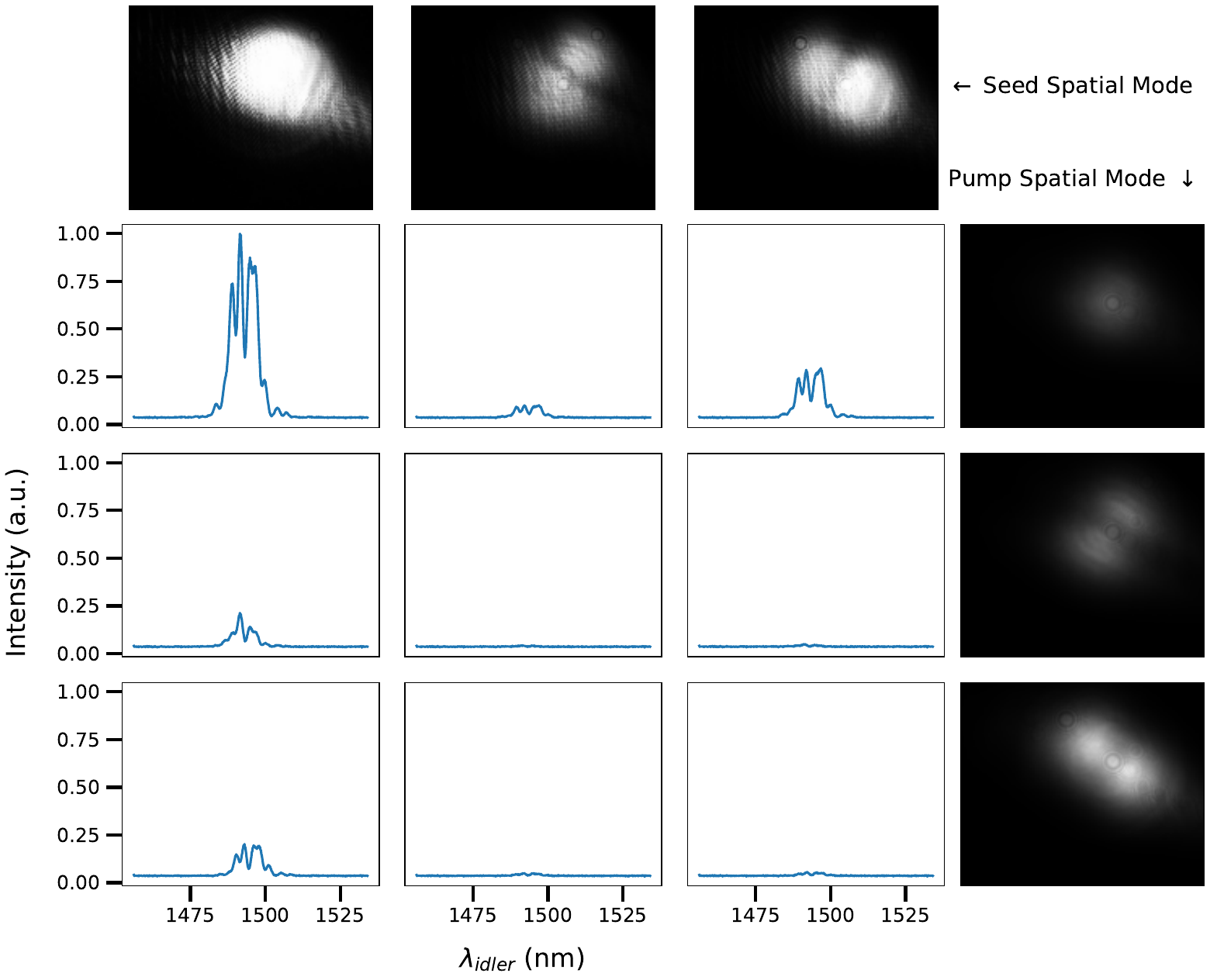}
     \caption{Spectral intensity of stimulated emission with different combinations of input spatial modes for the process 1. The seed  wavelength is set to 830 nm.}
     \label{fig:p1_spatial}
\end{figure}

\begin{figure}[htbp]
      \includegraphics[width=\linewidth]{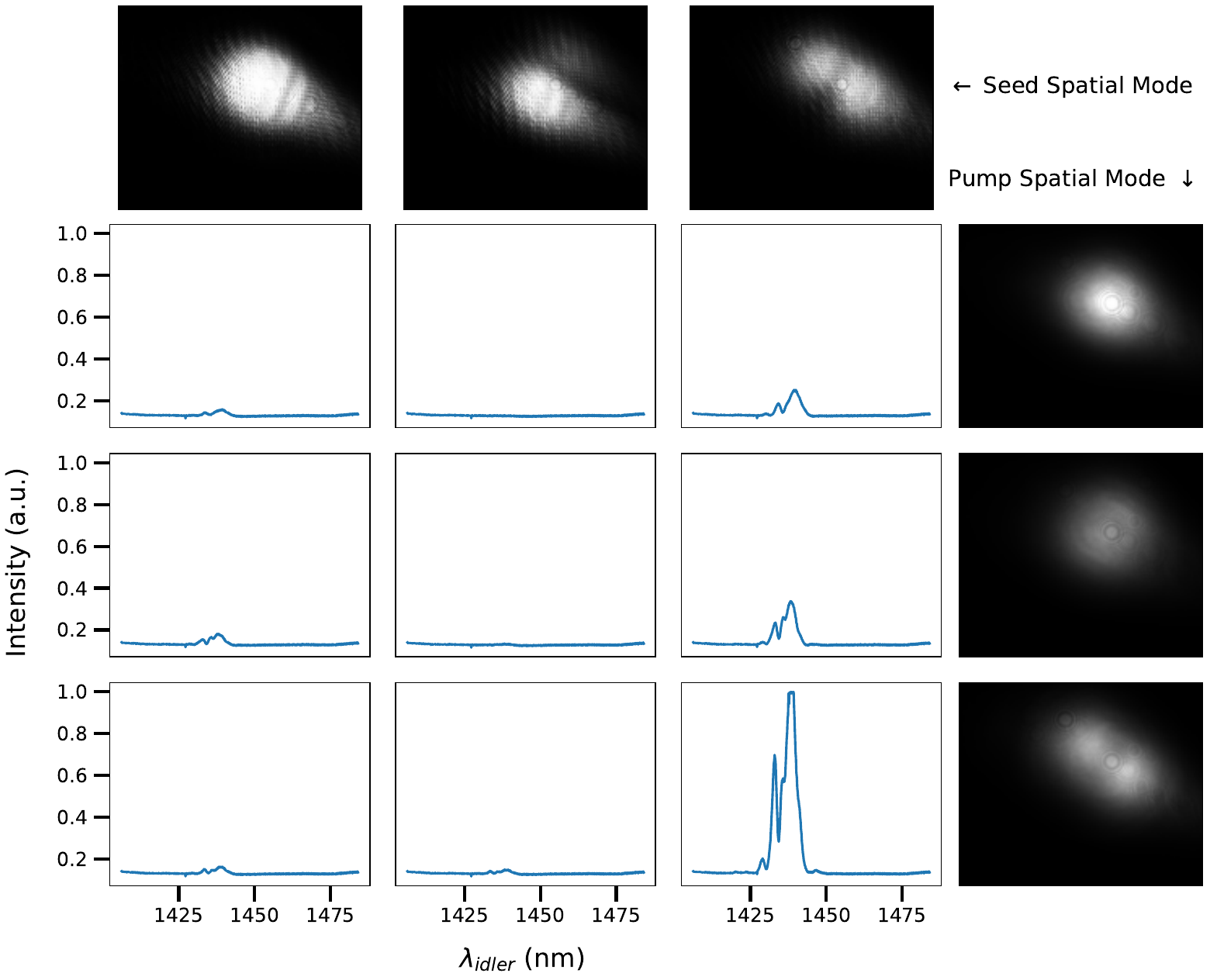}
     \caption{Spectral intensity of stimulated emission with different combinations of input spatial modes for the process 2. The seed wavelength is set to 850 nm.}
     \label{fig:p2_spatial}
\end{figure}

With the different spatial mode combinations, we can see that the process 1 occurs when the fundamental spatial mode is present for both the seed and the pump and the process 2 occurs when a higher-order spatial mode is present for both the seed and the pump.
We can see the results for process 1 in Fig. \ref{fig:p1_spatial} and the results for process 2 in Fig. \ref{fig:p2_spatial}

%\section{Heralding Efficiency}
%While we were measuring the $g^{(2)}$, we simultaneously kept track of the single count rate for both the signal and idler. From here we can calculate the heralding efficiency, which is about 15 \% for process 1 and 2 \% for process 2, as seen in Fig. \ref{fig:HE}. 

%\begin{figure}[htbp]
%      \includegraphics[width=\linewidth]{figures/figures_supplemental/HE_thin_ND_filter_both_process_20251010.pdf}
%     \caption{Heralding efficiency for both processes with respect to signal and idler singles counts.}
%     \label{fig:HE}
%\end{figure}

\section{Other PMF}
\begin{figure}[htbp]
      \includegraphics[width=\linewidth]{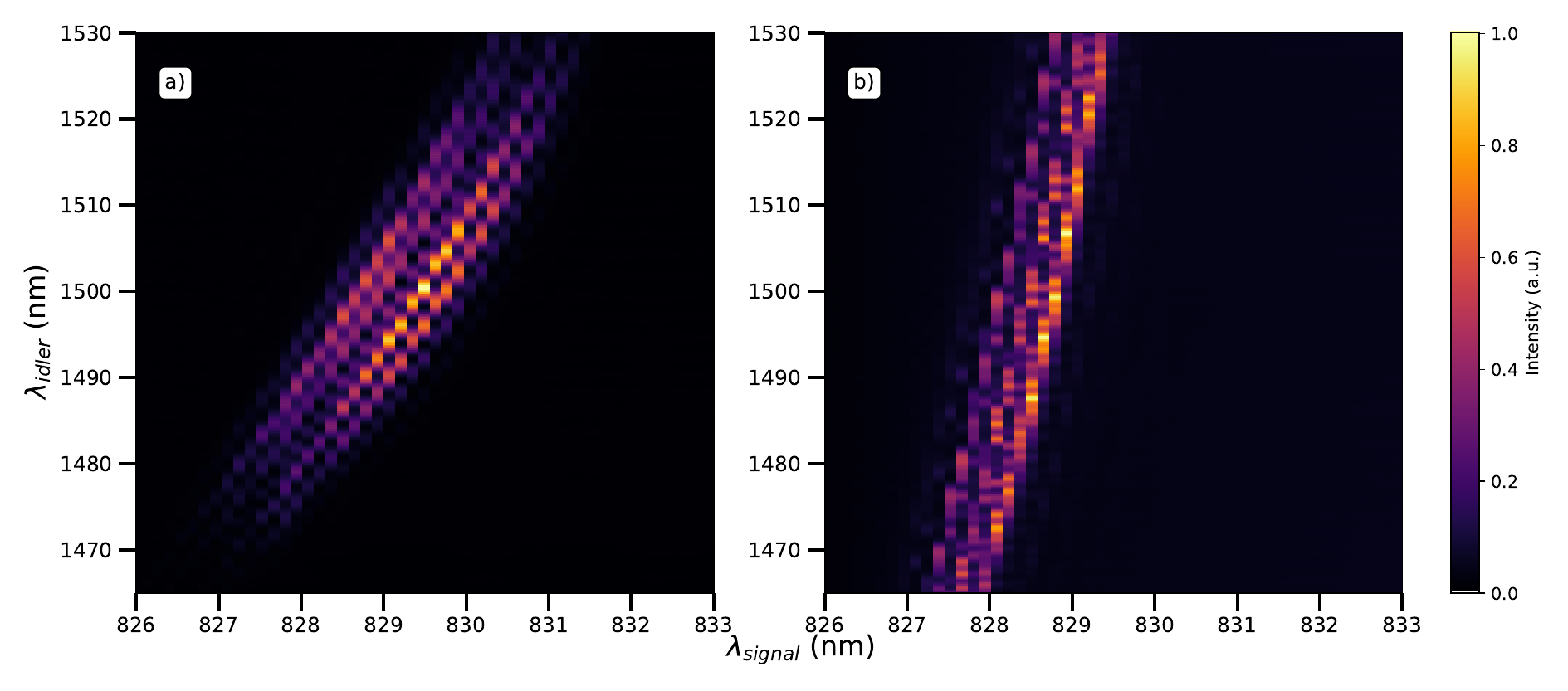}
     \caption{JSI via stimulated emission for non-high-extinction-ratio fibers of lengths (a) 1 meter and (b) 5 meter.}
     \label{fig:JSI_1m_5m}
\end{figure}

We have also tried other fiber lengths and types, specifically 1-meter- and 5-meter-long non-high extinction-ratio PMF. 
Here we show the JSIs of the process 1 for both of these fibers, as seen in Fig. \ref{fig:JSI_1m_5m}. Notice that the 5-meter JSI shows a thinner profile than the 1-meter one, as expected by a longer fiber's more restrictive phase matching condition~\cite{Garay-Palmett:23}.
We notice more streaks along the main JSI streak as compared to the high extinction-ratio PMF, so we use high extinction-ratio PMF for our source.

\section{Efficiency} \label{supp:eff}

We note that we have an efficiency of about 0.7 for the NIR light through the CWDM, and an efficiency of about 0.9 for the telecommunication light through the CWDM.
Efficiency for the NIR fiber coupling after free space filtering is about 0.75 and for the telecommunication fiber coupling after the filtering is also about 0.75.
We also note again the efficiency of the APD is 0.45 and the efficiency of the SNSPD is above 0.9.

From these numbers, we can see as we approach unit efficiency of these experimental components, we expect to see 226 kcps of coincidences from the source.

\end{document}